# Emergence of a coherent in-gap state in SmB$_6$ Kondo insulator revealed by scanning tunneling spectroscopy

Wei Ruan[1,2], Cun Ye[1,2], Minghua Guo[1,2], Fei Chen[3], Xianhui Chen[3], Guangming Zhang[1,2], and Yayu Wang[1,2] [†]

[1]*State Key Laboratory of Low Dimensional Quantum Physics, Department of Physics, Tsinghua University, Beijing 100084, P. R. China*

[2]*Collaborative Innovation Center of Quantum Matter, Beijing, China*

[3]*Hefei National Laboratory for Physical Science at Microscale and Department of Physics, University of Science and Technology of China, Hefei, Anhui 230026, P.R. China*

[†] Email: yayuwang@tsinghua.edu.cn

We use scanning tunneling microscopy to investigate the (001) surface of cleaved SmB$_6$ Kondo insulator. Variable temperature *dI/dV* spectroscopy up to 60 K reveals a gap-like density of state suppression around the Fermi level, which is due to the hybridization between the itinerant Sm 5*d* band and localized Sm 4*f* band. At temperatures below 40 K, a sharp coherence peak emerges within the hybridization gap near the lower gap edge. We propose that the in-gap resonance state is due to a collective excitation in magnetic origin with the presence of spin-orbital coupling and mixed valence fluctuations. These results shed new lights on the electronic structure evolution and transport anomaly in SmB$_6$.

The solution of the Kondo problem induced by single magnetic impurity embedded in a host metal is a major triumph of many body physics [1]. However, there are still many unresolved puzzles in the Kondo lattices formed by a dense array of magnetic ions coupled to itinerant electrons [2]. A particularly intriguing phenomenon concerns the $SmB_6$ Kondo insulator, which has a hybridization gap at the Fermi level ($E_F$) but the resistivity saturates at low temperature instead of diverging. Recent theories propose that $SmB_6$ is a $Z_2$ topological insulator (TI) with topologically protected surface state (SS) [3, 4], which naturally reconciles the dilemma. In contrast to the existing TIs that can be described by single electron band theory, this so-called topological Kondo insulator (TKI) represents the first TI in which many body effect is crucial. This new idea has inspired a host of experimental efforts searching for the topological SS. Although transport [5, 6], angle-resolved photoemission spectroscopy (ARPES) [7-11], cantilever magnetometry [12] and point contact spectroscopy [13] results all point to the existence of 2D electronic states, it has yet to be confirmed unambiguously if they are the topological SSs.

Another imperative task for understanding $SmB_6$ is to elucidate the evolution of its electronic structure. After all, at the heart of the TKI physics is the bulk band structure and topology. Moreover, it is quite generic for the Kondo lattice electronic structure to undergo a series of subtle reconstructions due to the intricate interaction between the local *f* electrons and itinerant bands [14, 15]. Even for the prototypical heavy fermion systems, there are still controversies regarding the nature of the electronic states [16, 17]. From this perspective, the $SmB_6$ Kondo insulator provides a unique opportunity for exploring the rich electron physics in the Kondo lattice system.

Owing to its ability to probe both the occupied and empty electronic states at the atomic scale, scanning tunneling microscopy (STM) has played a crucial role in unveiling the novel physics in both TIs and Kondo lattices. It is thus an ideal probe for clarifying the electronic structure evolution of the $SmB_6$ Kondo insulator and the existence of topological SS. In this paper, we report STM studies on the (001) surface of cleaved $SmB_6$. Scanning tunneling spectroscopy (STS) reveals the hybridization gap opening at high temperatures and the emergence of a low temperature resonance state within the gap. Implications of the STM results to the transport anomaly and topological SS in $SmB_6$ will be discussed.

High quality $SmB_6$ single crystals are grown by the Al flux method [18]. STM experiments are performed with a low temperature ($T$) ultrahigh vacuum system. The $SmB_6$ crystal is cleaved in the load-lock chamber with pressure better than $10^{-8}$ mbar at room $T$, and then immediately transferred to the STM stage cooled at 5 K. An electrochemically etched tungsten tip is used for the STM measurements. Before each measurement the tip is treated and calibrated carefully as described elsewhere [19]. Topographic images are scanned in constant current mode and $dI/dV$ (differential conductance) spectroscopy is measured by a lock-in amplifier with modulation frequency $f = 323$ Hz.

Figure 1a depicts the schematic cubic crystal structure of $SmB_6$. There are two possible cleaving planes parallel to the (001) surface, one between adjacent Sm/B planes and the other between two B planes within an octahedron. Fig. 1b shows the photograph of a cleaved (001) surface of $SmB_6$, which is flat and shining. All the STM data reported in this paper are taken on similar surfaces. Fig. 1c shows the typical electrical resistance ($R$) vs. $T$ curve measured on a $SmB_6$ crystal, which exhibits the characteristic insulating behavior at high $T$ and the saturation at $T < 5$ K (inset). The magnetic spin susceptibility also exhibits an anomaly at the low $T$ regime, as shown in Fig.1d.

Figures 2a to 2d display the STM images of four types of surface structures we have observed so far, and the insets are the corresponding Fourier transforms (FT). The first phase (Fig. 2a) has an atomically flat square lattice (denoted as "A1" phase), but the distance between two adjacent atoms is ~ 3 Å, or $a_0/\sqrt{2}$ ($a_0 = 4.1$ Å is the lattice constant of $SmB_6$). The only possible way to generate such a lattice constant is to cleave at the blue plane in Fig. 1a, and expose the Sm-B network schematically shown in Fig. 2e. In fact the topography in Fig. 2a already reveals a weak $\sqrt{2} \times \sqrt{2}$ superstructure, indicating the existence of two inequivalent sub-lattices. The second phase shown in Fig. 2b also exhibits a square lattice ("A2" phase), but with lattice constant $a_0$ formed by the B atoms from either the blue or purple cleaved plane in Fig. 1a. The dominant patterns in the third phase shown in Fig. 2c are donuts ("D1" phase) forming an incomplete square lattice, sitting on the periodic "A1" phase that is responsible for the FT spots. The atoms consisting the donuts are unknown, but most likely they are four B atoms self-assembled at the center of an "A1"-phase square, because the height of the donut layer is ~ 1 Å. The fourth phase shown in Fig. 2d also exhibits donuts ("D2" phase) patterns, but the donuts here have a larger diameter. Moreover, the underlying square lattice has a lattice constant equals $a_0 = 4.1$ Å, which can be seen from the FT in Fig. 2f or directly in the zoom-in image in Fig. 3a. The structure of this phase is well-defined. As illustrated in Fig. 2h, the donuts consist of eight boron atoms exposed by the cleavage within a boron octahedron, and they sit on a square lattice of the "A2" phase. Among these four types of surface structures, the "D2" phase is the majority phase. We notice that the surface morphologies here have differences from that reported by two other groups [20, 21]. The most likely cause is that in our work the SmB$_6$ crystal is cleaved at room $T$, whereas in the other two papers the crystals are cleaved at low $T$. Different cleaving temperatures will affect the surface relaxation process, and may even determine the specific layer where the system might fracture [22].

Figures 2i to 2l show bias voltage ($V$) dependent $dI/dV$, which is approximately proportional to the local electron DOS at energy $E = eV$, taken at $T = 5$ K on the four types of SmB$_6$ (001) surfaces. Despite the distinctly different surface morphologies, the $dI/dV$ spectra share some important common features. They all exhibit a well-defined energy gap around $E_F$ ($V = 0$), a sharp peak at negative energy $E_1 \sim -10$ meV, a broad hump centered around $E_2 \sim -160$ meV, and a large zero bias conductance (ZBC) corresponding to the residual DOS at $E_F$. The positions of the peak and hump features vary slightly for the four surfaces, but the overall $dI/dV$ lineshape is highly similar to each other. The energy gap and large ZBC are also similar to that observed in point contact spectroscopy on cleaved SmB$_6$ [13].

Given the mysterious $R$ vs. $T$ behavior in SmB$_6$, it is particularly informative to elucidate the $T$ evolution of electronic structure. We have performed $dI/dV$ spectroscopy over a wide $T$ range on donuts in the majority "D2" phase as shown in Fig. 3a. In Fig. 3b we show the large bias $dI/dV$ spectra taken at four representative $T$s. At the highest $T = 60$ K, an asymmetric gap-like feature already exists around $E_F$, so does the hump at $E_2 \sim -160$ meV. With decreasing $T$, the high energy hump feature remains nearly unchanged but the low energy electronic states change dramatically. In Fig. 3c we zoom into the low energy range with more detailed $T$ variations of the $dI/dV$ curve (black open circles). Upon cooling from 60 K, the DOS suppression around $E_F$ deepens continuously. At $T = 40$ K, the peak at $E_1 \sim -10$ meV starts to emerge. With further decrease of $T$, the peak becomes more pronounced and meanwhile a dip near -25 meV starts to develop. At the lowest $T = 5$ K, the spectrum exhibits a strongly asymmetric lineshape with the sharp peak-dip resonance located at the lower edge of the energy gap.

The most important question regarding the variable $T$ spectroscopy is the nature of the gap-like feature persisting to 60 K and the in-gap resonance emerging below 40 K. We propose that the high $T$ gap is formed by the hybridization between the itinerant Sm 5$d$ band and localized Sm 4$f$ band. The hybridization process originates from the Kondo screening of individual $Sm^{3+}$ $4f^5$ local moments by itinerant 5$d$ electrons, forming a narrow composite fermion band and a small energy gap around $E_F$. The Kondo screening scenario is supported by the magnetic susceptibility displayed in Fig. 1d, which shows a deviation from the high $T$ Curie-Weiss behavior at $\sim$ 120 K. Recent ARPES measurements also observed the hybridization gap opening at around 150 K [10].

To account for the tunneling through both the itinerant and localized bands in the Kondo lattice system, the $dI/dV$ curves on $SmB_6$ should be simulated by the co-tunneling model [23-27] instead of the Fano model for single magnetic impurity. The total tunneling conductance $G(eV)$ in this case can be expressed as:

$$G(eV) \propto \text{Im} \int d\omega \, f'(\omega - eV) \int_{B.Z.} d^3k [t_c \quad t_f] \begin{bmatrix} G_{cc}(\omega,k) & G_{cf}(\omega,k) \\ G_{fc}(\omega,k) & G_{ff}(\omega,k) \end{bmatrix} \begin{bmatrix} t_c \\ t_f \end{bmatrix}.$$

Here $G_{ij}(\omega,k)$ ($i, j=c, f$) are different components of the retarded Green's function, and $t_c/t_f$ are the tunneling probabilities of electrons through the itinerant electrons and localized

*f*-electrons respectively. Moreover, the thermal convolution effect at finite *T* is also accounted for [25]. As discussed in the supplementary materials (Part A), both the high energy hump feature at -160 mV and the low energy hybridization gap can be simulated very well by the co-tunneling model (the fit for the 60 K curve is shown in Fig. 4a). The energy levels of the two *f*-bands used in our simulation are also close to those probed by ARPES.

The sharp peak-dip feature of the in-gap resonance below 40 K, however, cannot be fit well by the co-tunneling model alone (Fig. S1 in supplementary). We propose that it represents a new collective mode unique to the $SmB_6$ Kondo insulator. The peak feature has totally different *T* evolution from the hybridization gap, and cannot be explained by simple thermal broadening effect either (Fig. S2). We further notice that neutron scattering [28], Raman spectroscopy [29] and optical conductivity [30] experiments on $SmB_6$ all detected a low energy magnetic resonance mode appearing at $E = 14$~$20$ meV below 40 K. Therefore, we propose that the sharp peak-dip feature observed in our *dI/dV* spectroscopy represents a new collective resonance mode unique to the $SmB_6$ Kondo insulator. Due to the small energy separation and the spatial variations, it is hard to distinguish this peak and the *f*-band in the ARPES results.

We found that the spectra below 40 K can be fit with a Gaussian peak when the hybridization gap background is subtracted. STS results on similar *f*-electron states have also been fit by Gaussian curves in previous reports [14, 15]. We have performed the co-tunneling plus Gaussian simulations to all *dI/dV* spectra (red lines in Fig. 3c), and the fit for the 5 K curve is shown in Fig. 4a. The *T* dependence of the peak spectral weight and full width at half maximum (FWHM) extracted from the fitting are shown in Fig. 4b, which clearly illustrate the emergence of the resonance below 40 K. Fig. 4c shows $T$ dependence of the extracted gap amplitude and quasiparticle scattering rate $\gamma_f$ of the $f$-electrons. The gap amplitude, hence the hybridization strength, gradually decreases with increasing $T$. If we simply extrapolate the gap amplitude linearly to high $T$, the gap will close at roughly 150 K, which is consistent with the ARPES results [10].

From the above analysis, in conjunction with the neutron and Raman results, we propose that this in-gap resonance can be understood based on the singlet wave function describing the Kondo screening of $Sm^{3+}$ $4f^5$ local moments and the valence mixing of $Sm^{2+}$ $4f^6$ non-magnetic ions. Due to the presence of interactions between different $Sm^{3+}$ sites and the spin-orbit coupling, the spin triplet state can be regarded as the possible low energy in-gap excitation [31]. This is consistent with the decrease of magnetic susceptibility starting from $T \sim 50$ K in Fig. 1d, which is caused by a more rapid suppression of DOS at $E_F$. Moreover, such a resonance in magnetic origin can strongly scatter the itinerant charge carriers, resulting in the dramatic increase of $R$ shown in Fig. 1c. In contrast, the much weaker insulating behavior at higher temperature is due to the slow DOS suppression by the hybridization process. Therefore, both the transport and magnetization anomalies can be understood consistently by the emergence of an in-gap resonance mode of magnetic origin below 40 K.

The last piece of puzzle is the saturation of resistivity at low $T$, and whether $SmB_6$ is a TKI. It is highly plausible that the large ZBC observed in our STS comes from the topological SS [3, 4], but direct STM evidence for TKI will be quasiparticle interference (QPI) patterns showing SS with helical spin texture, as demonstrated on the $Bi_2Te_3$ family TIs [32-34]. We have performed preliminary STM studies of the QPI phenomenon on the "A2" phase of $SmB_6$ (001) surface, which has relatively clean and flat surface (supplementary Part B). However, so far we have not detected any clear SS features in the DOS maps or its FT images. We emphasize that we cannot rule out the existence of topological SS based on this because several factors may prevent the observation of the expected QPI patterns. First, strong surface disorders may result in large scattering rate of the SS electrons, making the QPI phenomenon extremely weak. Second, the new resonance mode observed here, which resides in the hybridization gap thus may couple strongly with the topological SS, can also suppress the QPI features. Third, our STM study is performed at $T = 5$ K, when the saturation of resistivity is barely present. Therefore, future STM studies on $SmB_6$ surfaces with better morphology at $T$ well below 5 K are needed to detect the topological SS, if they do exist. This will be a challenging task due to the lack of an ideal cleaving plane in $SmB_6$. In fact, a recent ARPES study reveals polarity-driven surface states formed by B dangling bonds that is sensitive to surface conditions [35]. This topologically trivial metallic SS can explain both the resistivity saturation and the absence of QPI on a rough surface.

In summary, variable temperature STS on $SmB_6$ reveals the hybridization gap opening at temperatures above 60 K and the emergence of a collective in-gap resonance at $T < 40$ K. The electronic structure evolution revealed here can partly explain the anomalous transport and magnetic susceptibility behavior. Although some features of the spectroscopy are consistent with the existence of topological SS, no QPI pattern has been directly visualized. Therefore, it remains to be unambiguously proved if $SmB_6$ is a TKI.

We thank Peng Cai, Piers Coleman, Xi Dai, Yifeng Yang, and Xiaodong Zhou for helpful discussions. This work was supported by the National Natural Science Foundation and MOST of China (grant No. 2010CB923003, 2012CB922002). F. Chen and X.H. Chen acknowledge financial support from the 'Strategic Priority Research Program' of the Chinese Academy of Sciences (grant No. XDB04040100).

**Figure Captions:**

FIG. 1. (a) Schematic crystal structure of $SmB_6$ where two possible cleaving planes parallel to the (001) surface are shown in purple and blue. (b) Photograph of the cleaved (001) surface of a $SmB_6$ crystal. (c) The $R$ vs. T curve of $SmB_6$ shows the insulating behavior at high $T$ and the saturation at $T < 5$ K (inset). (d) Inverse magnetic susceptibility as a function of $T$ shows the deviation from the high $T$ Curie-Weiss behavior at ~ 120 K.

FIG. 2. (a)~(d) STM images on cleaved (001) surface of $SmB_6$. Four types of surface structures are observed, including atomically resolved "A1" (a) and "A2" phases (b), donut-like "D1" (c) and "D2" (d) phases. The weak $\sqrt{2} \times \sqrt{2}$ pattern in "A1" phase is indicated by the dashed circles in (a). The insets are the corresponding FTs. (e)~(h) Schematic top view of the four cleaved surfaces. The black circles in (g) and (h) mark the positions of the donuts. (i)~(l) $dI/dV$ spectra taken at $T = 5$ K on the representative locations marked by solid circles with corresponding colors in (a)~(d). Vertical offset is used for clarity.

FIG. 3. (a) High-resolution image of the majority "D2" phase. The STS measurements at different temperatures are all done on donuts in this phase. (b) High energy $dI/dV$ spectra taken on the donuts in the "D2" phase at four selected $T$s. (c) Low energy spectra (black open circles) taken at varied $T$s from 5 K to 60 K, revealing the emergence of the resonance peak below 40 K, and the simulations of the $dI/dV$ spectra (red lines). For curves above 40 K, the hybridization gap are simulated only by the co-tunneling model. For curves below 40 K, the data are simulated by the co-tunneling model and a Guassian term for the resonance peak.

FIG. 4. (a) Theoretical simulation of the $dI/dV$ spectra. For $T = 60$ K only the co-tunneling model is used to simulate the hybridization gap. For $T = 5$ K, the simulated gap feature (red curve) is subtracted from the raw data, and the remaining resonance peak is fitted by a

Gaussian term (blue curve). (b) The weight and FWHM of the Gaussian peak as a function of *T*, showing the disappearance of coherence peak above 40 K. (c) *T*-dependent gap amplitude and quasiparticle scattering rate $\gamma_f$ of *f*-electrons extracted from the co-tunneling simulation (Fig. 3c).

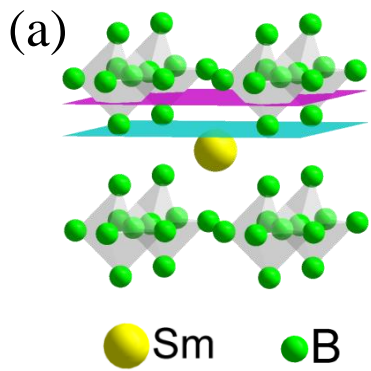
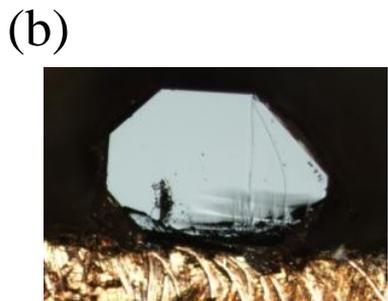
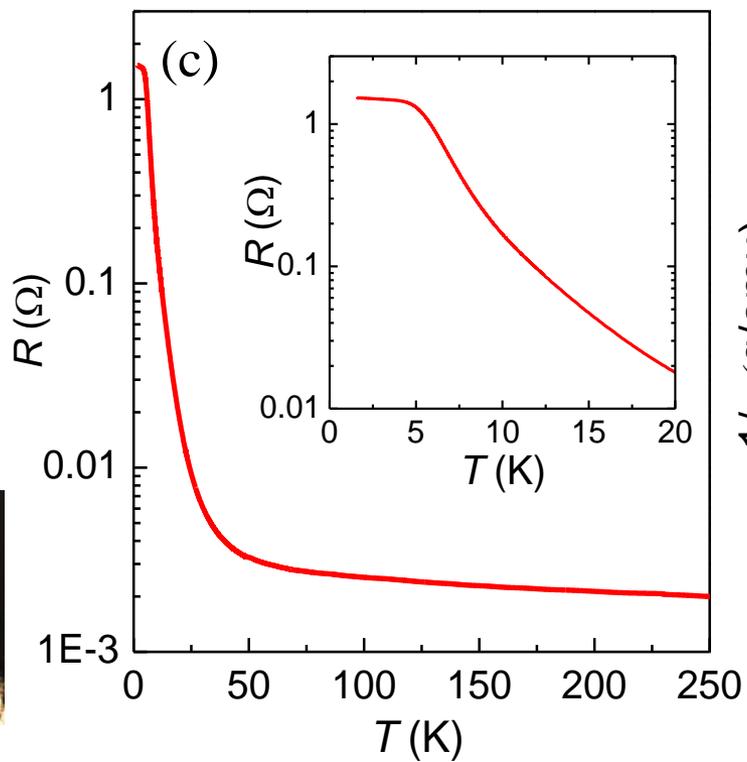
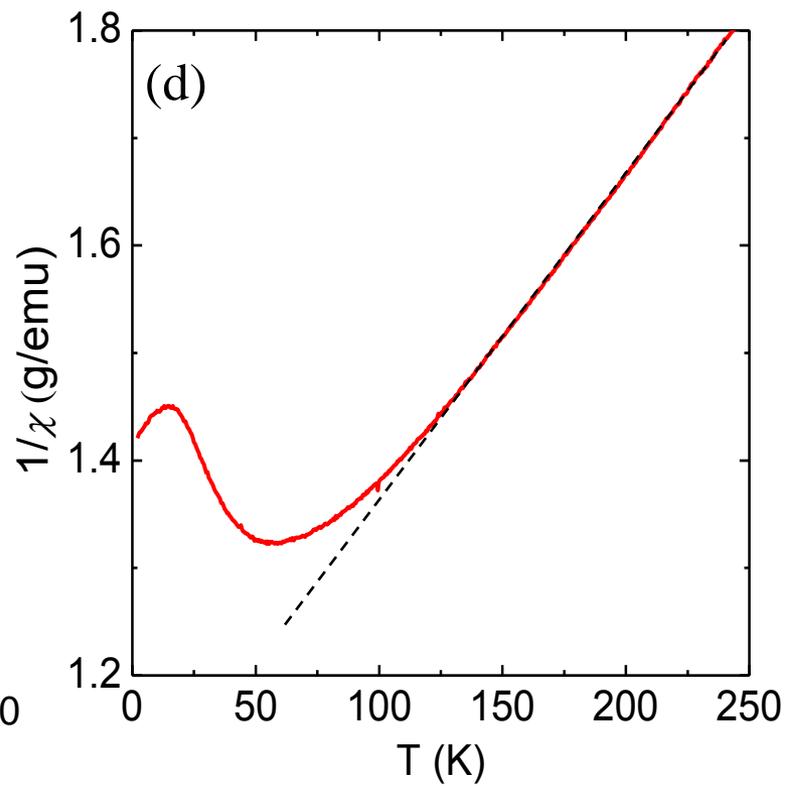

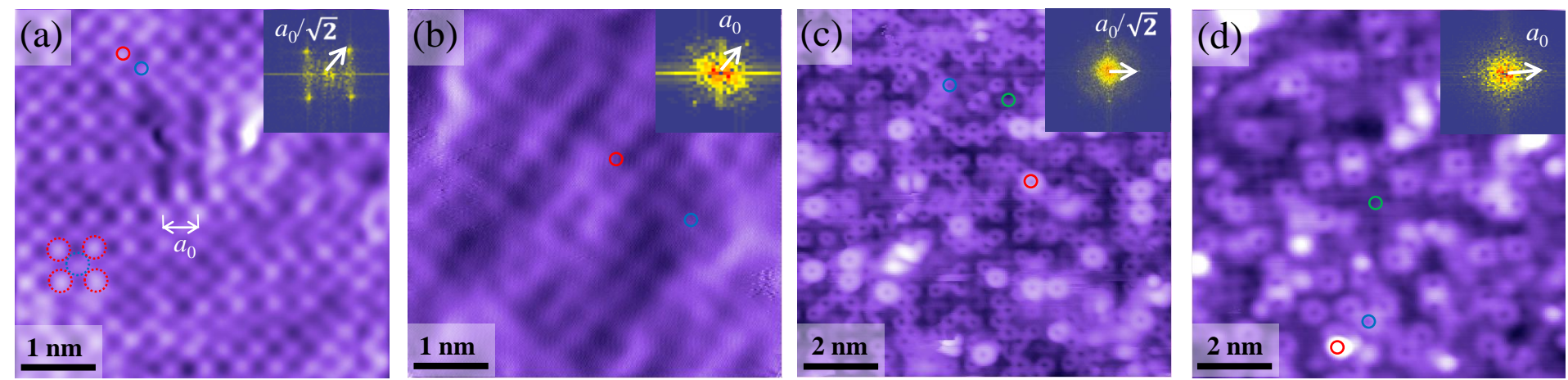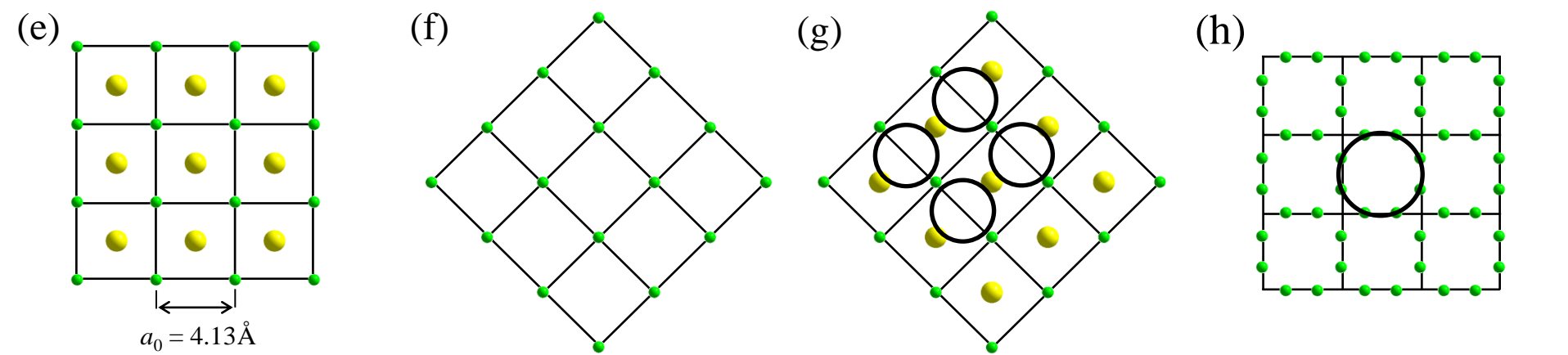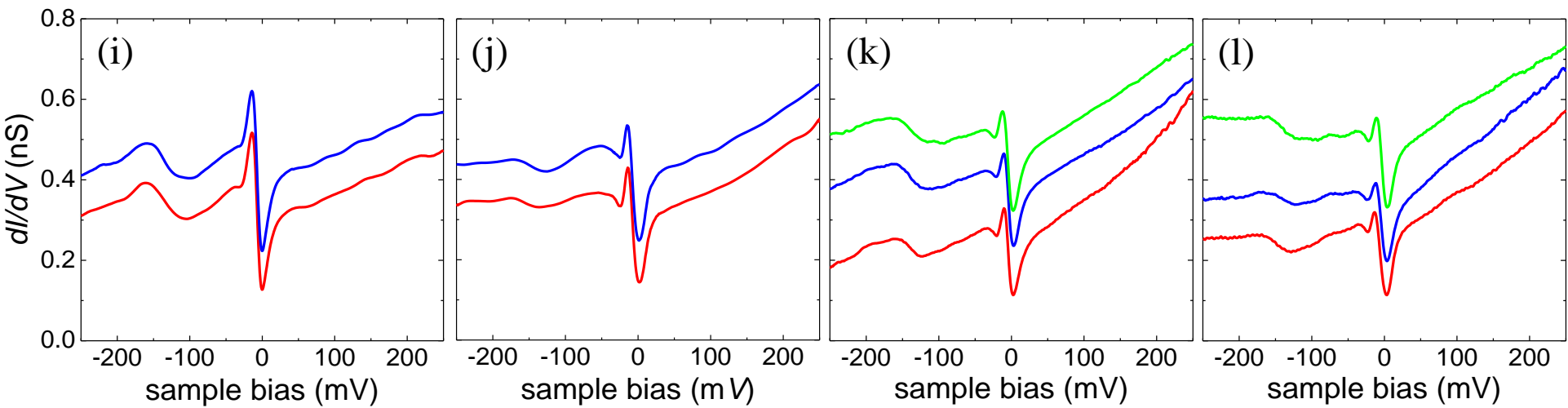

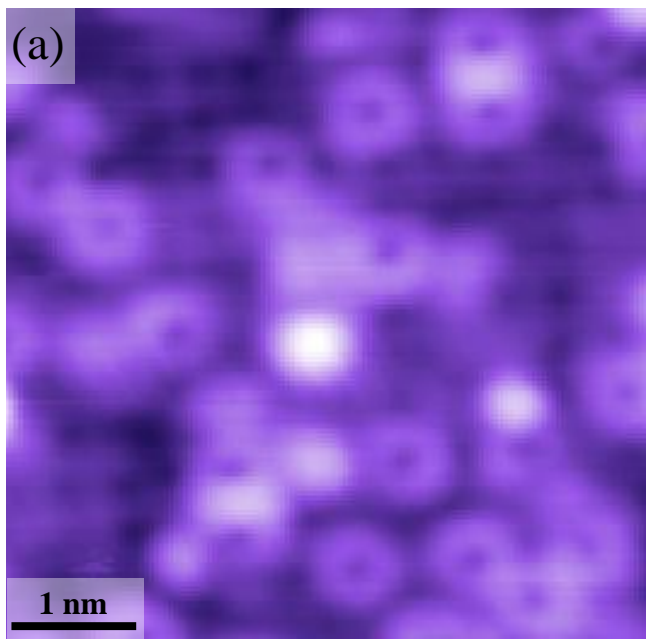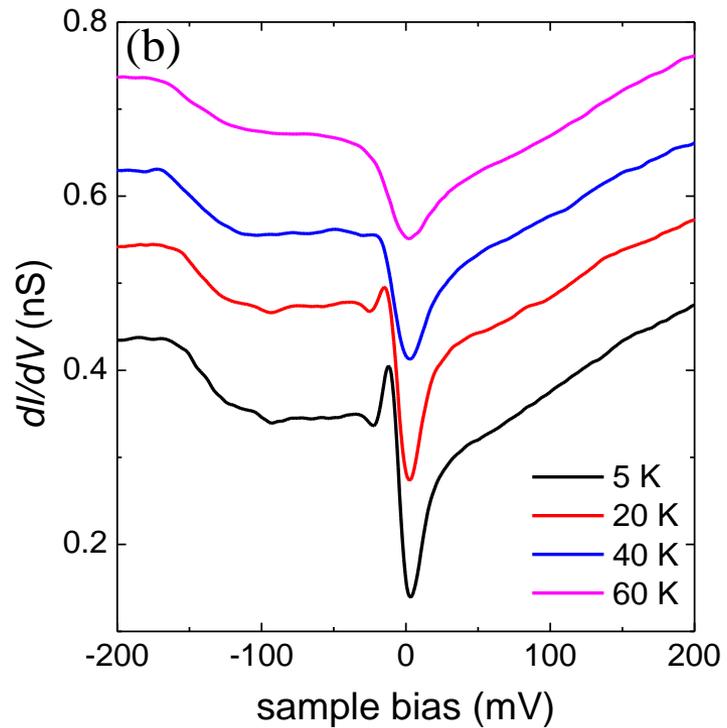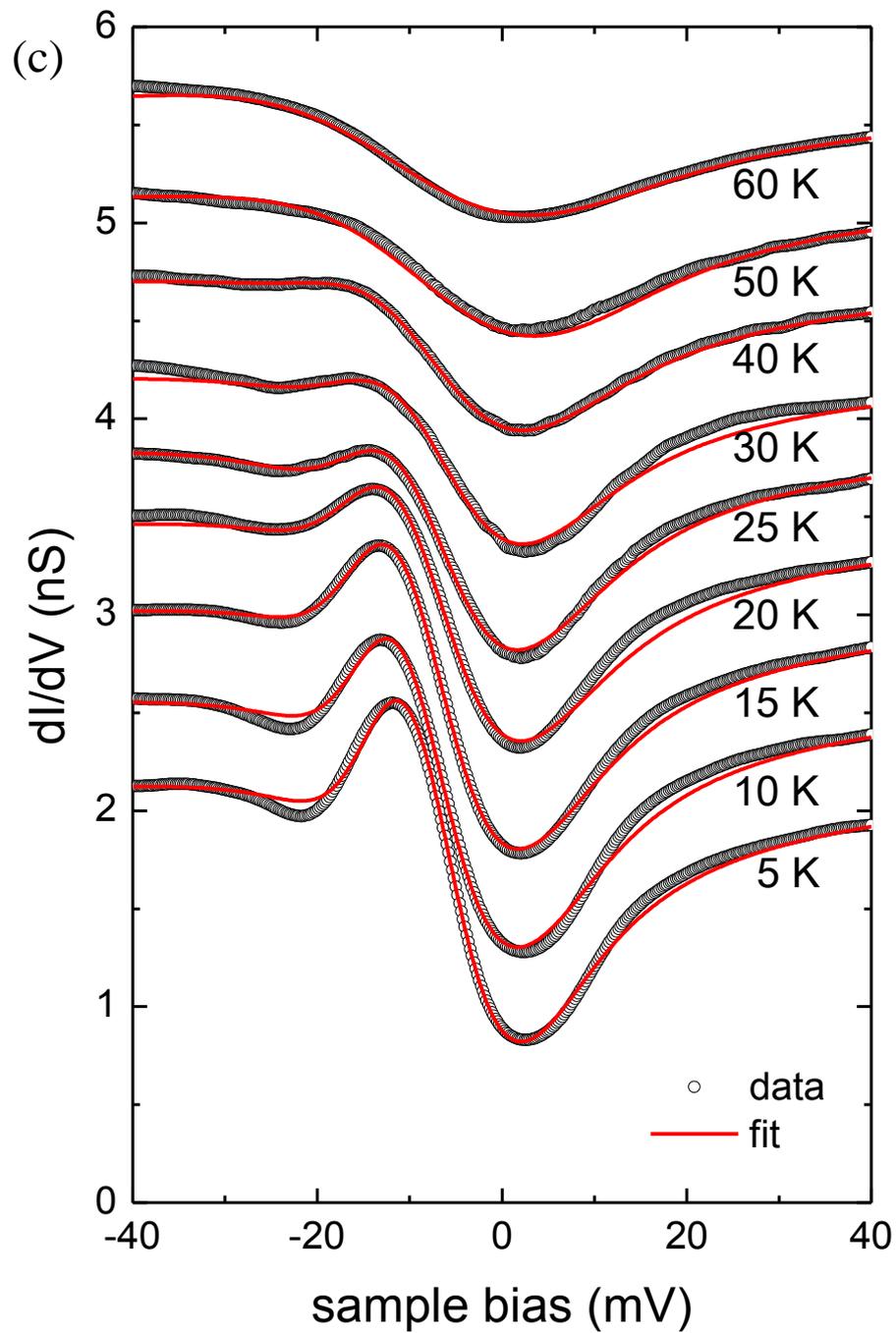

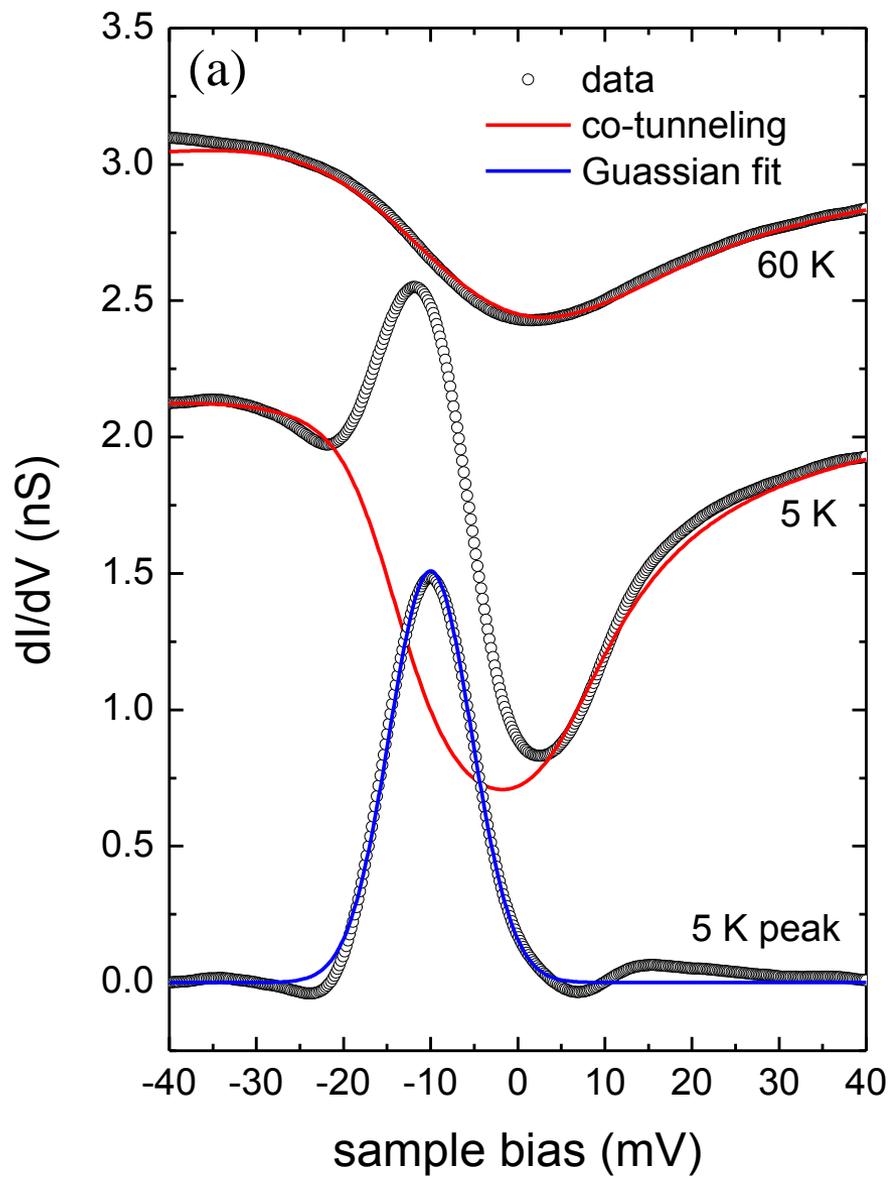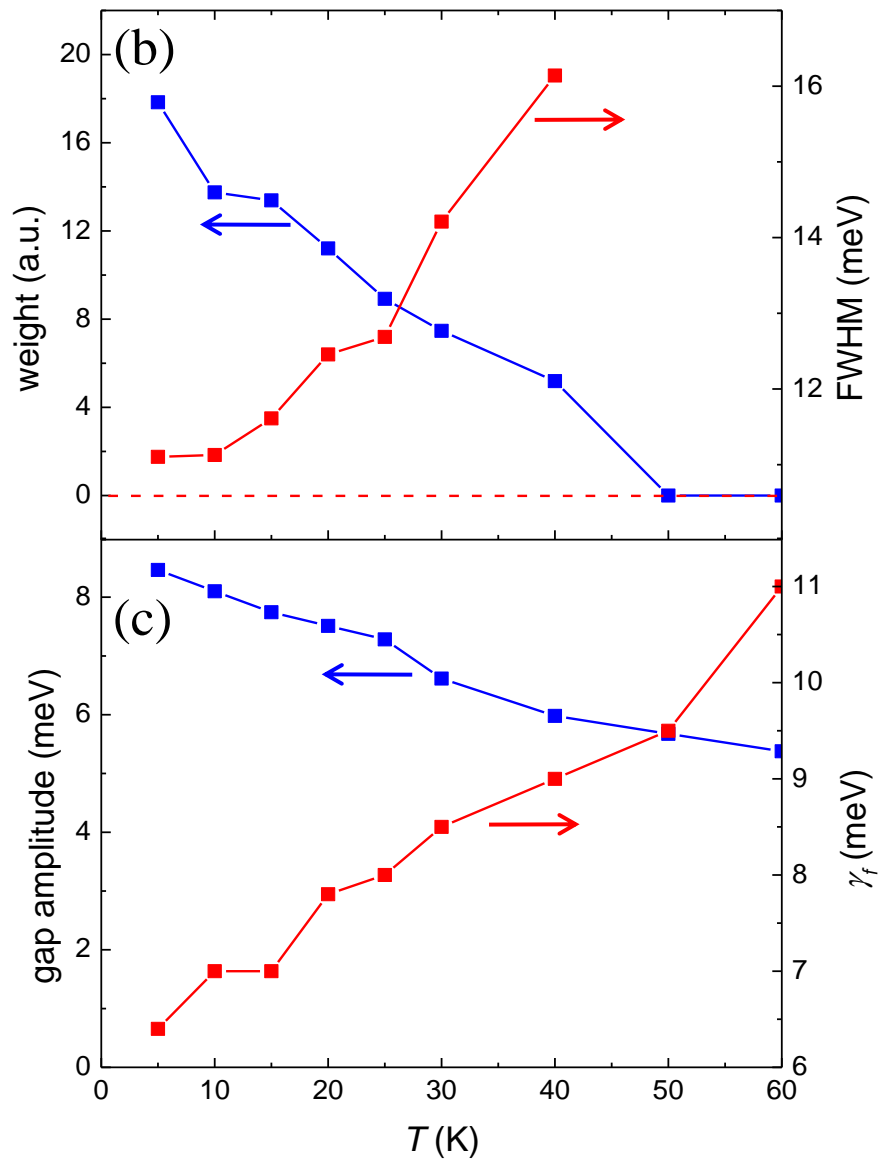

# Supplementary materials

# Emergence of a coherent in-gap state in $SmB_6$ Kondo insulator revealed by scanning tunneling spectroscopy

Wei Ruan[1,2], Cun Ye[1,2], Minghua Guo[1,2], Fei Chen[3], Xianhui Chen[3], Guangming Zhang[1,2], and Yayu Wang[1,2] [†]

[1]State Key Laboratory of Low Dimensional Quantum Physics, Department of Physics, Tsinghua University, Beijing 100084, P. R. China

[2]Collaborative Innovation Center of Quantum Matter, Beijing, China

[3]Hefei National Laboratory for Physical Science at Microscale and Department of Physics, University of Science and Technology of China, Hefei, Anhui 230026, P.R. China

[†] Email: yayuwang@tsinghua.edu.cn

**Contents:**

**A: The co-tunneling model simulation of the observed *dI/dV* spectroscopy**

**B: Preliminary quasiparticle interference (QPI) results**

Figure S1 to S6

References

## A. The co-tunneling model simulation of the observed *dI/dV* spectroscopy

In typical Kondo lattice systems involving both itinerant and localized electrons, the scanning tunneling spectroscopy should be simulated by using the co-tunneling model [1-4]. The total tunneling conductance $G(eV)$ can be expressed as:

$$G(eV) \propto \text{Im} \int d\omega\, f'(\omega - eV) \int_{B.Z.} d^3k [t_c \quad t_f] \begin{bmatrix} G_{cc}(\omega,k) & G_{cf}(\omega,k) \\ G_{fc}(\omega,k) & G_{ff}(\omega,k) \end{bmatrix} \begin{bmatrix} t_c \\ t_f \end{bmatrix}.$$

The tunneling Hamiltonian can be expressed as

$$H_T = \sum_k (t_c d_k^\dagger c_k + t_f d_k^\dagger f_k + \text{h.c.}).$$

Here $d$, $c$, and $f$ are annihilation operators of the tip, itinerant and localized electrons respectively. The general physics of Kondo lattice can be captured by the hybridization model (which can be derived from the periodic Anderson model) with Hamiltonian

$$H = \sum_k (\varepsilon_k c_k^\dagger c_k + \varepsilon_f f_k^\dagger f_k) + \sum_k (v_k c_k^\dagger f_k + \text{h.c.}).$$

Hence the Green's functions are given by

$$G_{cc}(\omega, k) = \frac{1}{G_{cc}^0(\omega,k)^{-1} - v_k^2 G_{ff}^0(\omega,k)},$$

$$G_{ff}(\omega, k) = \frac{1}{G_{ff}^0(\omega,k)^{-1} - v_k^2 G_{cc}^0(\omega,k)},$$

$$G_{cf}(\omega, k) = G_{cc}^0(\omega,k) v_k G_{ff}(\omega,k),$$

where the bare Green's functions are $G_{cc}^0(\omega,k) = \frac{1}{\omega - \varepsilon_k + i\gamma_c}$ and $G_{ff}^0(\omega,k) = \frac{1}{\omega - \varepsilon_f + i\gamma_f}$.

In this model, the hybridization gap can be estimated by second-order perturbation:

$$\Delta_h = 2 \frac{v^2}{D/2} = \frac{4v^2}{D},$$

where the hybridization strength takes a constant value $v_k = v$, and $D$ is the conduction band width.

ARPES studies on SmB$_6$ [5-7] have clearly resolved an electron-like conduction band centered at X point in the Brillouin zone coming from the 5$d$-orbital of Sm, and three flat 4$f$-bands of Sm at roughly -15 meV, -160 meV, and -960 meV respectively. In our simulation of the low energy $dI/dV$ spectroscopy near $E_F$, we model the conduction band by $\varepsilon_k = 2t(\cos \tilde{k}_x + \cos \tilde{k}_y + \cos \tilde{k}_z) + \mu$, and the $f$-band by $\varepsilon_f = 2\chi_0(\cos \tilde{k}_x + \cos \tilde{k}_y + \cos \tilde{k}_z) + 4\chi_1 \cos \tilde{k}_x \cos \tilde{k}_y \cos \tilde{k}_z + \varepsilon_f^0$.

The main parameters we used are $t$ = -750 meV, $\mu$ = 3000 meV, $\chi_0$ = 0.5 meV, $\chi_1$ = -0.02 meV, and $\varepsilon_f^0$ = -3 meV. $\tilde{k} = (\tilde{k}_x, \tilde{k}_y, \tilde{k}_z) = k - (\pi, 0, 0)$ is defined as the deviation from the X point. For simplicity here we use the isotropic conduction band instead of the observed elliptical band and with only one Fermi circle around the X point instead of three around X, Y, and Z point. We further simplify the model by taking the quasiparticle scattering rate as $\gamma_c = \gamma_f$ = 4 meV, and the hybridization strength $v$ = 100 meV. Because the conduction band width is around $D$ = 12|$t$| = 9 eV, the estimated hybridization gap is $\Delta_h$ = 4$v^2$/$D$ = 4.4 meV, in quantitative agreement with the experimental value.

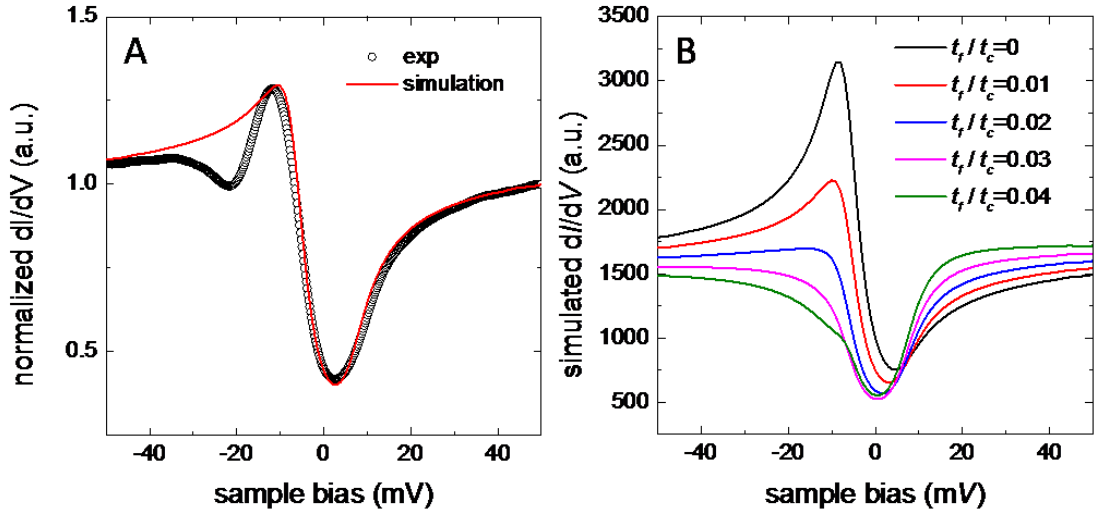

Fig. S1 (A) The experimental $dI/dV$ curve measured at 5 K and the simulated curve using the co-tunneling model with $t_f/t_c$ = 0.013. (B) Co-tunneling simulations using different values of $t_f/t_c$.

This simple model can capture the essential features if we only focus on the density of states near $E_F$. For example, it reproduces the hybridization gap at $T$ = 60 K very well, as shown in Fig. 3c and 4a in the main text. However, the sharp peak-dip peak that emerges below 40 K cannot be fit well by using this co-tunneling model alone. Shown in Fig. S1(A) is the best simulation of the observed curve at $T = 5$ K, in which the ratio $t_f/t_c$ is taken to be 0.013. It can be seen clearly that the dip at -20 meV cannot be captured by this fit at all. Fig. S1(B) displays the simulation results using different values of $t_f/t_c$, none of which reproduces the experimental curve.

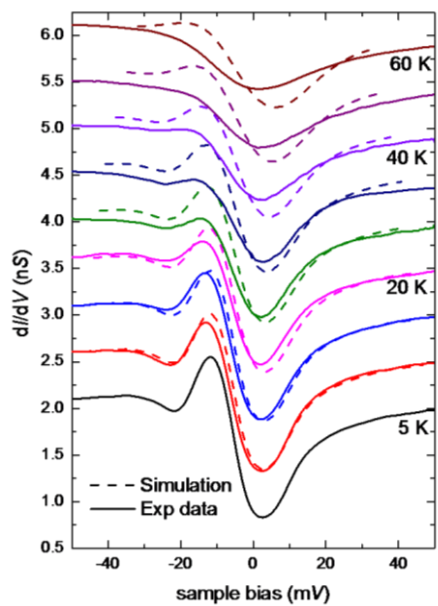

Fig. S2 Thermal broadening simulation at different temperatures. We take the $T = 5$ K curve and do thermal convolution for higher temperatures. The thermal effect is too weak to suppress the resonance peak feature up to 60 K.

In the experimental $dI/dV$ data, we clearly observe a resonance peak that disappears at $T > 40$ K, which is well below the onset temperature of the hybridization gap. To exclude the effect of thermal broadening, in Fig. S2 we take the $T = 5$ K spectrum and thermally convolve it at different temperatures (the broken lines). It is evident that the suppression of the peak cannot be accounted for by the thermal effect. Thus the peak emerging below 40 K represents a new electronic state. This point is supported by other experimental probes such as neutron scattering, Raman spectroscopy and optical conductivity which observed a magnetic excitation at similar energy and temperature scales.

Therefore, the peak and gap feature should be treated separately, as discussed in the main text. We use the co-tunneling model only to reproduce the hybridization gap and a Guassian term for the resonance peak, as shown in Fig. 4(a) in the main text. The main parameters we used here are $t = -750$ meV, $\mu = 3000$ meV, $\chi_0 = 0$ meV, $\chi_1 = 0$ meV, and $\varepsilon_f^0 = -6$ meV. We then perform the same procedure to all the spectra between 5 K and 60 K, and the simulated curves are shown in Fig. 3(c) of the main text. The simulation and fitting parameters are extracted and summarized in Fig. 4(b) and 4(c) of the main text.

To simulate the *dI/dV* spectroscopy for larger energy range, we further introduce a second flat *f*-band with $\chi_0' = 1$ meV, $\chi_1' = 0$, $\varepsilon_f^{0'} = -160$ meV, $\gamma_f' = 30$ meV, $v' = 100$ meV, and we assume this *f*-band does not interact with the one near $E_F$. The schematic electronic structure is illustrated in Fig. S3(A). The result of the simulation is shown in Fig. S3 (B), which can reproduce the hump feature at ~ -160 meV very well.

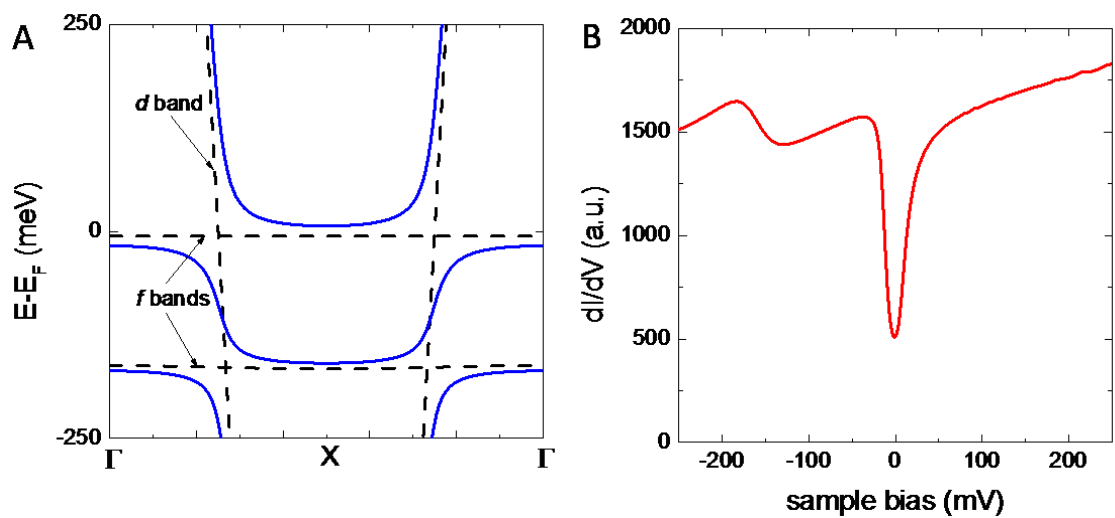

Fig. S3 (A) Illustration of the hybridization process. The itinerant *d*-band interacts with the two flat *f*-bands. (B) The simulated large energy range spectrum reproduces the hump feature at ~ -160 meV.

## B. Preliminary quasiparticle interference (QPI) results

To provide direct evidence for the existence of topological surface states, we have managed to perform *dI/dV* mapping on an area of 600 Å × 600 Å on the "A2" phase, whose topography is shown in Fig. S4(A). The surface looks smoother than the phases with donuts, but is still quite rough compared to the surface of the $Bi_2Te_3$ family TIs. The spatially averaged *dI/dV* spectrum taken in this area is shown in Fig. S4(B).

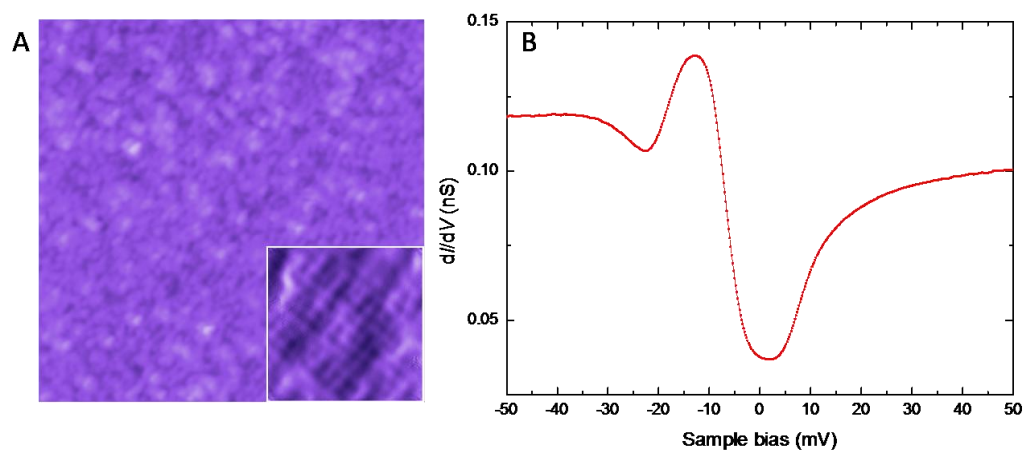

Fig. S4 (A) A 600 Å × 600 Å topography on the "A2" phase. The inset shows a 50 Å × 50 Å area with atomic resolution, where the atoms form a 4 Å × 4 Å square lattice. (B) The spatially averaged *dI/dV* spectrum in this area.

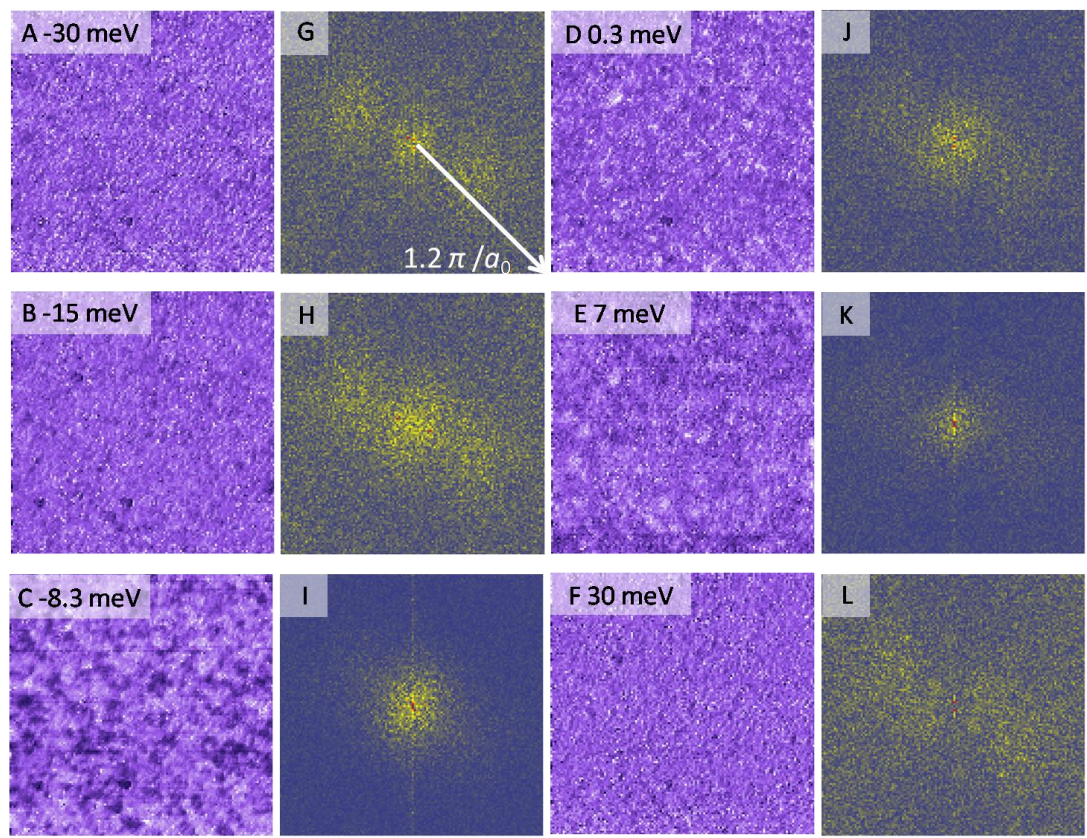

Fig. S5 (A)~(F) The *dI/dV* map at several representative bias voltages. (G)~(L) The corresponding FT of each *dI/dV* map.

Figure S5 displays several *dI/dV* maps at representative bias voltages from -30 meV to 30 meV covering the gap and peak features. Each map consists of 128 ×128 pixel numbers. The strongest spatial variation of the DOS occurs at energy near the resonance peak, as displayed in Fig. S5 (C) with bias voltage at -8.3 mV. However, no obvious wave-like QPI patterns as that observed in the $Bi_2Te_3$ family TIs are observed here.

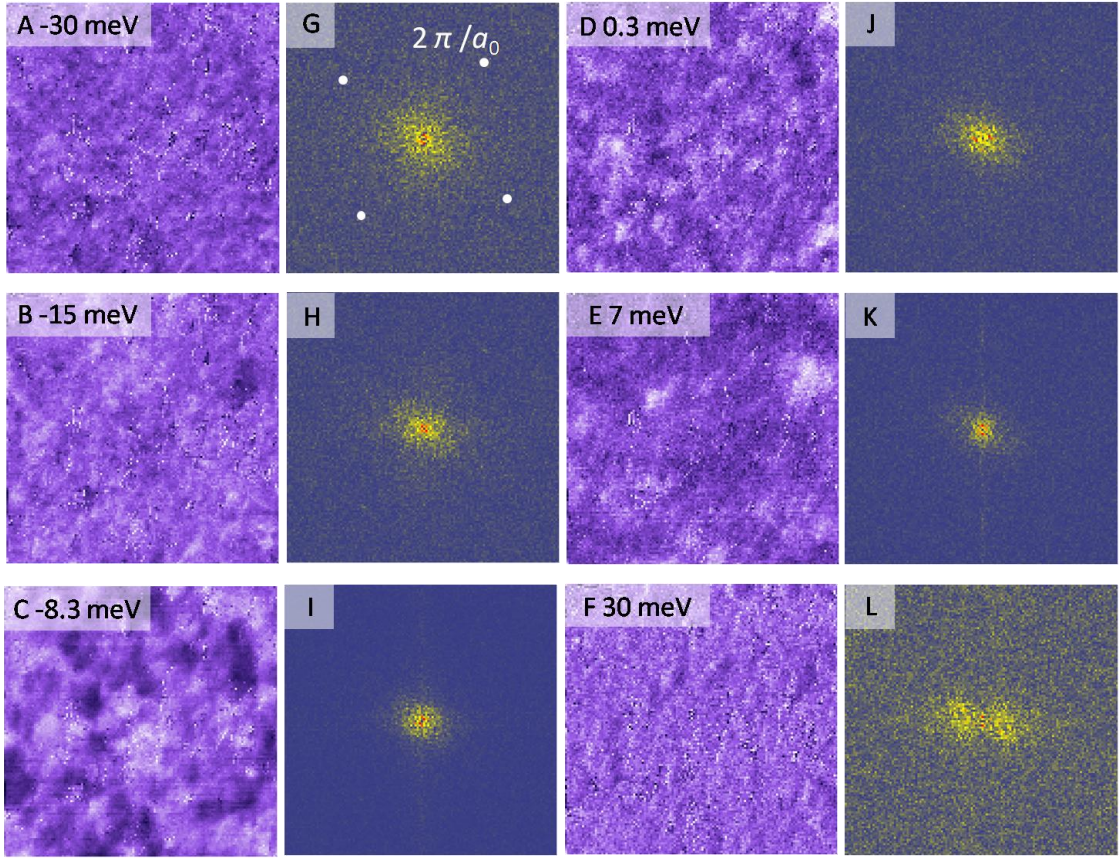

Fig. S6 (A)~(F) The *dI/dV* map at several representative bias voltages on another 200 Å × 200 Å area. (G)~(L) The corresponding FT of each *dI/dV* map. The small white dots indicate the Bragg points of the lattice.

Figure S6 shows the *dI/dV* mapping done on another 200 Å area with the same pixel number, which covers a larger k-space. Again, we do not see any clear signature of QPI patterns in the whole Brillouin zone. As discussed in the main text, the fact that we do not observe the expected QPI on the $SmB_6$ (001) surface cannot exclude the possibility that it is a topological Kondo insulator. Future STM studies to much lower temperatures on $SmB_6$ with better surface quality are needed to detect the QPI caused by the topological SSs, if they do exist.